\documentclass[pre,twocolumn,showpacs,preprintnumbers,amsmath,amssymb]{revtex4}

\usepackage{graphicx}
\usepackage{dcolumn}
\usepackage{bm}

\begin{document}

\newcommand{\bea}{\begin{eqnarray}}
\newcommand{\eea}{  \end{eqnarray}}
\newcommand{\bit}{\begin{itemize}}
\newcommand{\eit}{  \end{itemize}}

\newcommand{\be}{\begin{equation}}
\newcommand{\ee}{\end{equation}}
\newcommand{\ra}{\rangle}
\newcommand{\la}{\langle}
\newcommand{\U}{\widetilde{U}}


\def\bra#1{{\langle#1|}}
\def\ket#1{{|#1\rangle}}
\def\bracket#1#2{{\langle#1|#2\rangle}}
\def\inner#1#2{{\langle#1|#2\rangle}}
\def\expect#1{{\langle#1\rangle}}
\def\e{{\rm e}}
\def\proj{{\hat{\cal P}}}
\def\tr{{\rm Tr}}
\def\H{{\hat H}}
\def\Hdag{{\hat H}^\dagger}
\def\Lop{{\cal L}}
\def\Ehat{{\hat E}}
\def\Edag{{\hat E}^\dagger}
\def\Shat{\hat{S}}
\def\Sdag{{\hat S}^\dagger}
\def\Ahat{{\hat A}}
\def\Adag{{\hat A}^\dagger}
\def\U{{\hat U}}
\def\Udag{{\hat U}^\dagger}
\def\Zhat{{\hat Z}}
\def\Phat{{\hat P}}
\def\Op{{\hat O}}
\def\id{{\hat I}}
\def\x{{\hat x}}
\def\P{{\hat P}}
\def\Px{\proj_x}
\def\Pr{\proj_{R}}
\def\Pl{\proj_{L}}


\title{Spectral behavior of contractive noise}

\author{Gabriel G. Carlo}
\affiliation{Departamento de F\'\i sica, CNEA, Libertador 8250, (C1429BNP) Buenos Aires, Argentina}
\author{Alejandro M. F. Rivas}
\affiliation{Departamento de F\'\i sica, CNEA, Libertador 8250,
(C1429BNP) Buenos Aires, Argentina}
\author{Mar\'\i a E. Spina}
\affiliation{Departamento de F\'\i sica, CNEA, Libertador 8250,
(C1429BNP) Buenos Aires, Argentina}

\email{spina@tandar.cnea.gov.ar,rivas@tandar.cnea.gov.ar,carlo@tandar.cnea.gov.ar}

\date{\today}


\begin{abstract}

We study the behavior of the spectra corresponding to quantum
systems subjected to a contractive noise, i.e. the environment
reduces the accessible phase space of the system, but the total
probability is conserved. We find that the number of long lived
resonances grows as a power law in $\hbar$ but surprisingly there
is no relationship between the exponent of this power law and the
fractal dimension of the corresponding classical attractor. This
is in disagreement with the predictions of the fractal Weyl law
which has been established for open systems where the probability
is lost under the effect of a projective noise.

\end{abstract}

\maketitle

\section{Introduction}
\label{sec1}

In recent years there has been a great interest in the study of
open quantum systems. The motivation
for this upsurge can be found in the recent development of quantum
computation and information \cite{Nielsen,Preskill} (even though
some related aspects have been under investigation from long ago,
like in scattering theory, for example). The noise coming
from the environment has always been the major drawback for any
realistic attempt to implement a quantum computer and a serious
obstacle for quantum information purposes. At the same time we
have realized that we know much less about quantum open systems
than about their closed counterparts. Therefore, their study has
become an extremely active topic in fundamental physics \cite{Weiss}.

It is in this context and for the case of scattering systems that
the fractal Weyl law for the number of long-lived resonances has
been conjectured and tested by means of several examples
\cite{Nonnenmacher}. The classical invariant distribution in these
cases, i.e. the fractal hyperbolic set of all the trajectories
non-escaping in the past and future (the repeller), plays a
fundamental role with respect to the quantum spectrum of
quasibound states (the resonances). Indeed, the aforementioned law
says that \cite{conjecture} the number of resonances that decay at
the slowest pace (the long-lived ones) grows as
$\hbar^{-(1+d_H)}$, where $d_H$ is the partial Hausdorff dimension
of the repeller. We can trace back these studies to the proposal
and proof of a fractal Weyl upper bound for a Hamiltonian flow
showing a fractal trapped set \cite{Sjostrand}. Later, results
that seem to strongly support the validity of the conjectured law
have been obtained for different kind of systems ranging from
smooth to hard walls potentials \cite{Hamiltonians}. But it is in
open quantum maps that the fractal Weyl law was more easily
tested \cite{qmaps}. In these systems, paradigmatic in the quantum
chaos literature, the resonances have been found to grow as
$\hbar^{-d}$, where $d$ is the partial fractal dimension of the
repeller. However, explorations of the spectral behavior of
quantum systems having different kinds of invariant classical
distributions associated with them are very scarce. In this sense
it is very interesting to ask what happens in the case of a
non-projective kind of noise, i.e., a contractive environment,
for example. Here we will focus on dissipative quantum operations,
whose action can be interpreted as a phase space contraction leading to dissipative
dynamics \cite{contrac}. There were very few attempts to study
the spectral properties of these kind of systems in previous
works. The only antecedent we were able to find in the literature
is the paper by Ermann et al. \cite{Ermann}. But in this case only
the spectrum of a discretized Perron-Frobenius operator was
considered. For this approximation to the classical problem (i.e. not
a proper quantization of it), the authors
found that the behavior of the corresponding long-lived resonances
follow the fractal Weyl law.

In this work we analyze the dissipative baker map. To obtain the
quantum counterpart we have implemented a standard procedure
in which  the noise superoperator is written in terms of
appropriately defined Kraus operators \cite{Kraus}. For this
model, all classical initial conditions fall on a strange
attractor after a few iterations. At the quantum level, this is
represented by a resonance with eigenvalue $1$, and the rest of
the spectrum lying inside the unit circle. In sharp contrast to
what has been found for the discretized classical dynamics, the
number of long-lived resonances of the superoperator grows as a
power law in $\hbar$, but the exponent is rather insensitive to
the dimension of the fractal invariant set.

This paper is organized as follows: in Section \ref{sec2} We present the definition and details
of the model we have used to study this kind of spectrum. In Section \ref{sec3} the numerical
results are analyzed and we present an interpretation of them supported by further
studies. Finally, we draw the conclusions in Section \ref{sec4}.

\section{Model system}
\label{sec2}

One of the most simple models one can think of at the time of studying
complex systems are chaotic maps. Regardless of their simplicity they
capture all the essential features of chaotic behavior and their quantization
allows to extend these advantages to quantum mechanics. All this turned
them into paradigmatic models for quantum chaos and the theory of dissipative systems
\cite{Haake,Ozorio 1994,Hannay 1980,Espositi 2005}.
We have investigated the spectral behavior of the dissipative baker map.
The classical map is defined on the 2-torus $\mathcal T^{2}=[0,1)$ x $[0,1)$ by
\begin{equation}
\mathcal B(q,p)=\left\{
  \begin{array}{lc}
  (2q,\epsilon \: p/2) & \mbox{if } 0\leq q<1/2 \\
  (2q-1,(\epsilon \: p+1)/2) & \mbox{if } 1/2\leq q<1\\
  \end{array}\right.
\label{classicalbaker}
\end{equation}
This transformation is an area-contracting, piecewise-linear map.
The map contracts the torus in the $p$ direction by a $\epsilon$
factor, stretches the unit square by a factor of two in the $q$
direction, squeezes it by the same factor in the $p$ direction,
and then stacks the right half onto the left one. After a few
steps any initial distribution falls into a fractal set, the
strange attractor.

In an even $N$-dimensional Hilbert space (where $N=(2 \pi
\hbar)^{-1}$), the quantum baker map is defined in terms of the
discrete Fourier transform in the momentum representation as
\cite{Saraceno1,Saraceno2}
\begin{equation}
\label{quantumbaker}
 B_{N}= \left(\begin{array}{cc}
  G_{N/2} & 0 \\
  0 & G_{N/2}\\
  \end{array} \right)G_{N}^{-1},
\end{equation}
with $$(G_{N})_{jk}=\dfrac{1}{\sqrt{N}} \exp\{-2\pi i(j+1/2)(k+1/2)/N\}.$$
$B_{N}$ is a unitary matrix and represents the quantum dynamics of the closed baker map.

As discussed in \cite{contrac} quantum dissipative processes can
be described by non-unital quantum operations. In this work the
dissipative noise is implemented by an $ N^2 \times N^2 $ Kraus
superoperator of the form:

\begin{equation}
M=\sum_{\mu = 0}^{N-1} A^{\mu}\otimes A^{\mu \dag}
\end{equation}

where

\begin{equation}
A^{\mu}=\sum_{i=\mu}^{N-1} \sqrt{\left( \begin{array}{c} i\\
                          i- \mu
                          \end{array}
\right) \epsilon^{i-\mu} (1-\epsilon)^{\mu}} \ket{i-\mu} \bra{i}
\end{equation}
are operators accounting for transitions towards the momentum
state $\ket{i=0}$. The coupling constant $ \epsilon $ coincides
with the dissipation parameter of the corresponding classical map.
M is a trace preserving ($\sum_{\mu} A^{\dag}_{\mu}A_{\mu} = 1 $)
and non-unital ($\sum_{\mu} A_{\mu} A_{\mu}^{\dag} \neq 1 $)
superoperator, which describes a process in which phase space
volume is not preserved. Superoperators of this type are obtained
from the integration of master equations derived from modelling a
microscopic interaction of a system (an oscillator, a large spin,
etc) with a thermal bath at zero temperature representing the
environment \cite{Liu,Dittrich,Braun}.

To describe the noisy evolution of the density matrix of the
system we compose M with the unitary map (\ref{quantumbaker}),
according to:
\begin{equation}
\label{superoper}
\$= (B_N\otimes B_N^{\dag}) \circ M.
\end{equation}

\section{Results}
\label{sec3}

\begin{figure}[h]
\begin{center}
\includegraphics*[width=0.9\linewidth,angle=0]{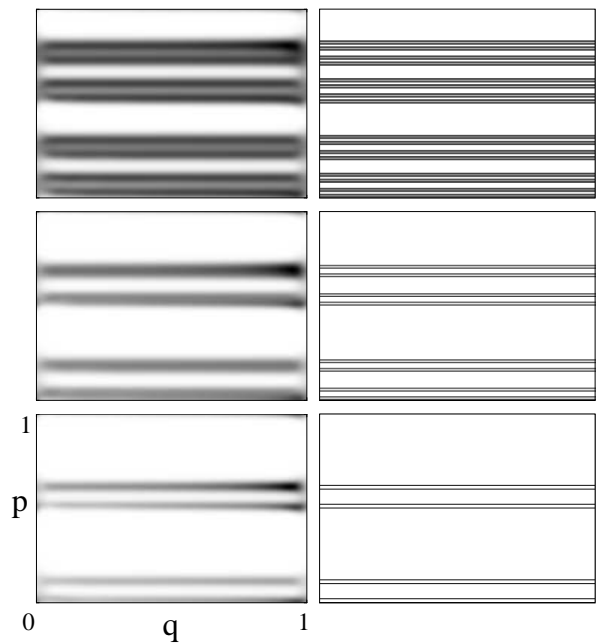}
\caption{Husimi representation of the invariant state of the
quantum dissipative baker map for (from top to bottom) $\epsilon=
0.8, 0.6, 0.4$ and $N=180$ (left column), and the corresponding
classical attractors (right column).
\label{f1}}
\end{center}
\end{figure}

The spectrum obtained by the diagonalization of the superoperator
of Eq.(\ref{superoper}) consists of a leading eigenvalue $\lambda=
1$ , and $N^2 - 1$ complex eigenvalues $\lambda$ (with $|\lambda|=
\exp ({-\gamma \over 2})$) inside the unit circle. Due to its
non-normality, $ \$ $ has distinct left $ \psi_{\lambda}^L $ and
right eigenoperators $ \psi_{\lambda}^R $ corresponding to each
eigenvalue. As expected, the Husimi function of the invariant
state $ \psi_0^R $ having eigenvalue $\lambda=1$, closely follows
the fractal structure of the attractor corresponding to the
classical dissipative map. This is shown in Fig. \ref{f1}, for
several values of $\epsilon$. In the classical case the
dissipation parameter $\epsilon$ determines the dimension of the
attractor $d=1+(\ln(2)/(\ln(2)-\ln(\epsilon))$ \cite{Gaspard}. In
the quantum case $\epsilon$ is related to the quantum phase space
contraction rate \cite{contrac}. We have verified that as $N$
increases, the phase space representation of $ \psi_0^R $ reveals
finer details of the attractor , reflecting the quantum to
classical correspondence.

\begin{figure}[h]
\begin{center}
\includegraphics*[width=1.\linewidth,angle=0]{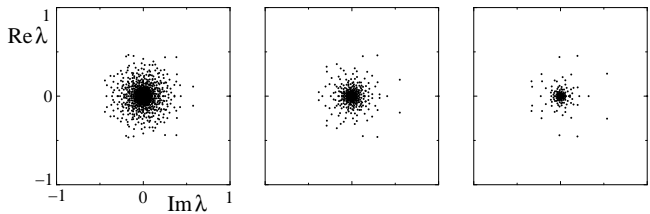}
\caption{Eigenvalues of the quantum dissipative baker map $ \$ $
in the complex plane for three different values of the
dissipation. From left to right: $ \epsilon=0.8,0.6,0.4$. In all
cases $N=180$.
\label{f2}}
\end{center}
\end{figure}

As a first approach to analyzing the behavior of the resonances,
in Fig. \ref{f2} we show different spectra of $\$$ for three
values of the dissipation parameter $\epsilon=0.8,0.6,0.4$ and a
Hilbert space dimension $N=180$. We notice that the longest-lived
eigenvalues, though change their position in the complex plane,
more or less keep their moduli while the radius $
r_{\lambda}$ of the dense circle where most of the eigenvalues
concentrate strongly shrinks for increasing values of the
dissipation. However, this radius cannot be directly related to
the parameter $\epsilon$, in contrast to what is observed in
\cite{Ermann} for the spectra of the discretized Perron-Frobenius
operator. This gives to the quantum version a seemingly more contractive
character in spectral terms.

\begin{figure}[h]
\begin{center}
\includegraphics*[width=1.\linewidth,angle=0]{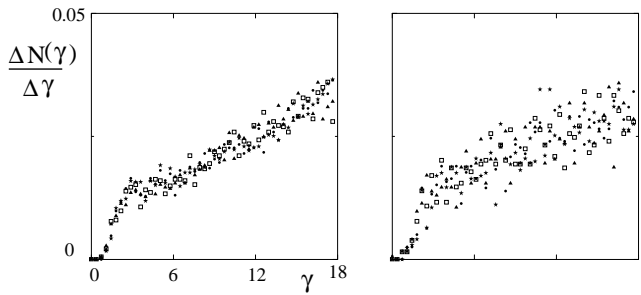}
\caption{Density of states $\Delta N(\gamma) \over \Delta \gamma$
as a function of $\gamma$ for different values of $N$
($90(\blacktriangle),100(\square),150(\bullet),180 (\star))$and
two values of the dissipation $\epsilon=0.8$ left and
$\epsilon=0.6$ right. The densities are normalized according to
$\int_0^{18}{\Delta N(\gamma) \over \Delta \gamma} =1$.
\label{f3}}
\end{center}
\end{figure}

To better show this point the differential radial density
distributions defined as $\Delta N(\gamma)\over \Delta \gamma$ are
plotted in Fig. \ref{f3}, for  different values of $N$. These
densities, normalized according to $\int_0^{18} {\Delta N(\gamma)
\over \Delta \gamma} = 1$, are practically N-independent for the
dimensions we are considering. They present a relative maximum at
a decay rate $\gamma \ll - 2 \ln \epsilon $, and then they
smoothly increase at larger values of $\gamma$. These profiles are
very different from the ones obtained in \cite{Ermann} for the
spectrum of the Perron-Frobenius operator, which peak around
$\gamma = -2 \ln \epsilon $, that is, the global relaxation rate
to the strange attractor, and decay rapidly for larger $\gamma$.

\begin{figure}[h]
\begin{center}
\includegraphics*[width=.8\linewidth,angle=0]{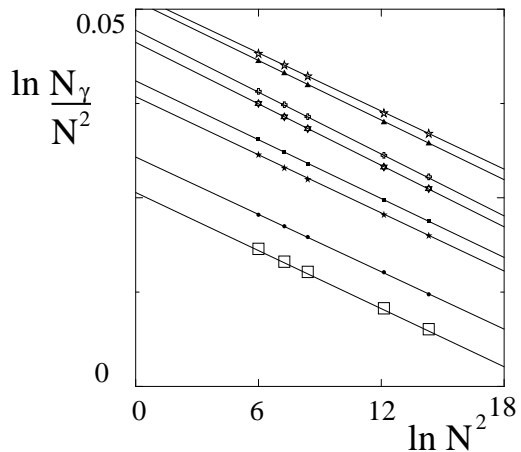}
\caption{Log-log plot for the fraction of long-lived resonances (with
$\gamma < \gamma_{\rm cut}$) as
a function of $N^2$ for $ \epsilon=0.8 $ and eight values of the
cut-off value ranging from $4 \leq \gamma_{cut} \leq 20 $. The
fitted slopes are in the interval $ 0.76 \leq \beta \leq 0.82$.
\label{f4}}
\end{center}

\end{figure}

We now come to the central point of this paper. We want to
see how the differences between a full quantization and a
discretization procedure of phase space which are suggested by the
comparison of the corresponding spectra affect the adherence to
the fractal Weyl law. Fig. \ref{f4} displays the fraction of long-lived
resonances $ {N_{\gamma<\gamma_{cut}} \over N^2}$ as a
function of $ N^2 $ for $ \epsilon= 0.8 $ and values of
$\gamma_{cut}$ ranging from $4$ to $20$. It can be clearly seen
that this fraction scales as $ {N_{\gamma<\gamma_{cut}} \over N^2}
\propto (N^2)^{-\beta}$ with an exponent $\beta \sim 0.8$ fairly
independent of the cut-off value $\gamma_{cut}$ used to
distinguish between the long-lived states and the short-lived
ones.
The existence of a power law dependence for $
{N_{\gamma<\gamma_{cut}} \over N^2}$ and the insensitivity of the
exponent to the choice of the cut-off value $\gamma_{cut}$ are
fundamental properties of the fractal Weyl law which has been
originally conjectured for projective noises. However, in the case
of our contractive noise the exponent which fits the numerical
results cannot be related in a direct way to the dimension of the
classical attractor. This dimension is equal to $ d=1.756$  for
$\epsilon=0.8$ and thus the relation $ \beta=2-d $ does not hold.

\begin{figure}[h]
\begin{center}
\includegraphics*[width=1.\linewidth,angle=0]{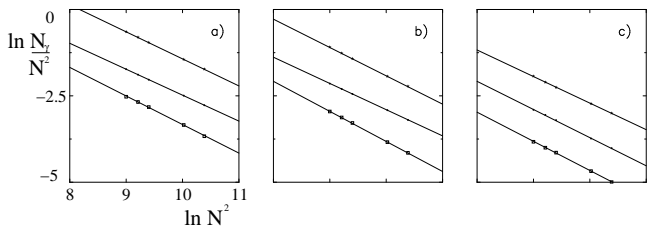}
\caption{Log-log plot for the fraction of long-lived resonances (with
$\gamma < \gamma_{\rm cut}$) as
a function of $N^2$ for  $
\epsilon=0.8({\star}),0.6({\bullet}),0.4({\square})$, with a)
$\gamma_{cut}=20 $, b) $\gamma_{cut}=15.2 $, c) $\gamma_{cut}=9.2
$ . The fitted slopes are $\beta=0.83,0.75,0.78$ for a),
$\beta=0.87,0.76,0.82$ for b) and $\beta=0.84,0.82,0.77$ for c).
\label{f5}}
\end{center}
\end{figure}

Moreover, the exponent $ \beta $  turns out to be practically
insensitive to the value of the dissipation parameter $ \epsilon$.
This remarkable fact can be appreciated in Fig. \ref{f5}  where $
{N_{\gamma<\gamma_{cut}} \over N^2}(N^2)$ is shown for different
values of the coupling constant $\epsilon=0.8,0.6,0.4 $ (and three
choices of $\gamma_{cut}$). In all the cases the data follow a
power law, with $\beta \sim 0.8 $, independently of the fractal
dimension of the corresponding classical attractor ( $ d=1.756$
for $\epsilon=0.8$, $d=1.576$ for $\epsilon=0.6$, $ d=1.431$  for
$\epsilon=0.4 $).

For comparison we have applied the superoperator formalism to the
well-studied case of an open Baker map subjected to a projective
noise modelled by a non-unitary operator $U_{open}$, by defining $
\$_{open}=U_{open}\otimes U_{open}^{\dag}$ . We verified that, as
expected, the fraction of long-lived resonances of $ \$_{open}$
scales with $N^2$ as does the fraction of resonances of $U_{open}$
with $N$ (we have taken $N$ values in the same range than for the
contractive case), thus following the fractal Weyl law.

\begin{figure}[h]
\begin{center}
\includegraphics*[width=1.\linewidth,angle=0]{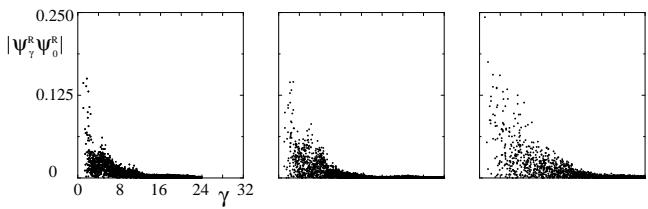}
\caption{Overlap $ |(\psi_{\lambda}^R \psi_0^R )| $ as a function
of $ \gamma $ for different values of the dissipation parameter
$\epsilon=0.8,0.6,0.4$ (from left to right) and $N=180$.
\label{f6}}
\end{center}

\end{figure}

\begin{figure}[h]
\begin{center}
\includegraphics*[width=0.9\linewidth,angle=0]{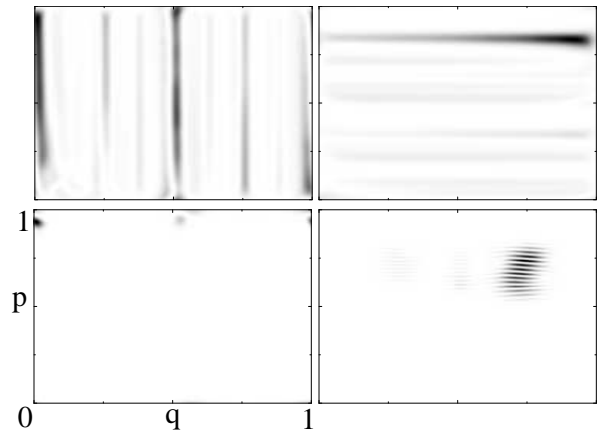}
\caption{Husimi representations of the a) left and b) right
eigenstates for the closest to the unit circle excited state
$|\lambda|=0.5845$  (upper panel) and for a state with a large
decay rate $|\lambda|=9.278 \times 10^{-5}$ (lower panel), $
\epsilon=0.8, N=180$. \label{f7}}
\end{center}
\end{figure}

In order to shed more light on these results we turn to analyze
the eigenvectors. To characterize the eigenstates of $ \$ $ we
consider the overlap of the right eigenstates with the invariant
state, by defining the measure $ |(\psi_{\lambda}^R \psi_0^R )| =
|Tr(\psi_{\lambda}^{R \dag} \psi_0^R)|$. Fig. \ref{f6} displays
the dependence of this measure with respect to $\gamma$, for
different values of $\epsilon$. It is clear that the overlap is in
the average larger for slow-decaying states, and decreases as the
decay rate increases.  But it is always small, this giving further
support to the strong contraction of the spectrum we have
observed.
The distinction between slow- and fast-decaying states is also
noticeable if we look at their distribution in phase space. For
this, we compute the Husimi representation of the right and left
eigenstates of $ \$ $, $ \bra{z}\psi_{\lambda}^{R,L}\ket{z}=
Tr(\psi_{\lambda}^{R,L \dag},\ket{z}\bra{z}) $ where $\ket{z}$ are
coherent states centered at $z=(q,p)$. As shown in Fig. \ref{f7} for a
typical slow-decaying state(upper panel) the Husimi density of the
right eigenstate reasonably follows the structure of the attractor
(though being much more localized than the invariant state),
while the left one has a delocalized pattern. For a state with a
large decay rate (lower panel), the probability pattern becomes
difficult to associate to the attractor.

\begin{figure}[h]
\begin{center}
\includegraphics*[width=0.7\linewidth,angle=0]{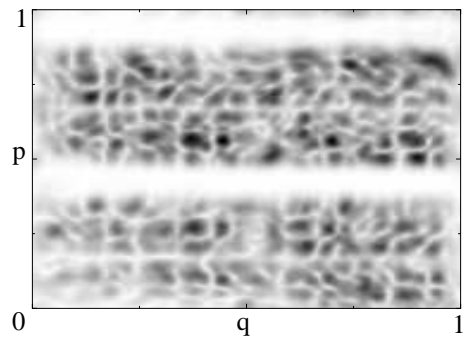}
\caption{Husimi representation for the projector of Eq.
(\ref{projector}), with $\lambda_{cut}=0.1348$ (or
$\gamma_{cut}=4$) , $ \epsilon=0.8, N=180$.
 \label{f8}}
\end{center}
\end{figure}

Although the long-lived eigenfunctions are morphologically
different from the invariant state, we have verified that they
have support on the phase space region corresponding to the
classical attractor by making use of a recently developed
representation especially suited for open systems \cite{Ermann2}.
We have built the sum
\begin{equation}
\sum_{\lambda=\lambda_{cut}}^{1}{\bra{z}\psi_{\lambda}^{R}\ket{z}
\bra{z} \psi_{\lambda}^{L}\ket{z} \over
\bracket{\psi_{\lambda}^{L}} {\psi_{\lambda}^{R}}}
\label{projector}
\end{equation}
and obtained the distribution shown in Fig. \ref{f8}, indicating a
clear localization of the long-lived states on the attractor
region in the semiclassical limit.

In view of these results we have found that though the
classical support of the long-lived resonances is the classical
attractor, their number scales with $N$ at a different pace than that
predicted by the fractal Weyl law.

\section{Conclusions}
\label{sec4}

We have studied the spectra of a paradigmatic model of the quantum
chaos and dissipative systems theories, the baker map with
dissipation, by following a standard quantization procedure based
on the Kraus representation of superoperators.  In contrast to
what happens for the discretized Perron-Frobenius operator in
\cite{Ermann}, we have found that the standard fractal Weyl law
does not hold in this case. Even if the fraction of long-lived
resonances does scale with the dimension following a power law
which is roughly insensitive to the cut off value of the decay
rate considered for the statistics, the exponent (which is
approximately constant) is not directly related to the fractal
dimension of the classical attractor.

In order to give an interpretation for this intriguing result we
have investigated the morphology of the eigenstates of the system.
In particular we have built the Husimi representation of the
projector constituted by the eigenfunctions with slow escape rate
\cite{Ermann2}. We found that its density concentrates on the
attractor region, strongly suggesting that one should expect, as
for the projective case, a connection between the statistics of
these long-lived resonances and the fractal dimension of the
strange attractor.

A possible cause for the lack of such natural connection might be
in the non-orthogonality of the eigenfunctions localized on the
strange attractor. In fact, the fractal Weyl law supposes the
quasi-orthogonality of the eigenfunctions supported by the fractal
set, since it is based on a Planck-cell partitioning of this set.
This works in the case of projective noises, but might not be the
case for contractive ones. This is an open problem and we are
currently studying \cite{future} the degree of non-orthogonality
induced by a contractive dynamics as an explanation for the
apparent non validity of the standard Weyl law revealed by the
present numerical investigation.

\section*{Acknowledgments}

Support from CONICET is gratefully acknowledged.

\vspace{3pc}


\end{document}